\documentclass[12pt,preprint]{aastex}
\begin{document}
\title{Measuring Feedback Using the Intergalactic Medium State and
Evolution Inferred from the Soft X-ray Background}
\author{Zhang, Pengjie\footnote{\it email: zhangpj@cita.utoronto.ca}}
\affil{Department of Astronomy \& Astrophysics, University of Toronto,
Toronto, ON M5S 3H8, Canada} 
\author{Pen, Ue-Li\footnote{\it email: pen@cita.utoronto.ca}}
\affil{Canadian Institute for Theoretical Astrophysics, Univ. of
Toronto, Toronto, ON M5S 3H8, Canada}
\begin{abstract}
We explore the intergalactic medium (IGM) as a potential source of the
unresolved soft X-ray background (XRB) and the feasibility to extract
the IGM state and 
evolution from XRB observations. We build two analytical models, the
continuum field 
model and the halo model, to calculate the IGM XRB mean flux, angular
auto correlation and cross correlation with galaxies. Our results 
suggest that the IGM may contribute a significant fraction to the unresolved
soft XRB flux and correlations. We calibrated non-Gaussian errors
estimated against our $512^3$ moving mesh hydro simulation and
estimate that the ROSAT all sky survey 
plus Sloan galaxy photometric redshift survey would allow a $\sim 10\%$
accuracy in the IGM XRB-galaxy cross correlation power spectrum measurement for
$800<l<5000$ and a $\sim 20\%$ accuracy in the 
redshift resolved X-ray emissivity-galaxy cross correlation power
spectrum measurement for $z\lesssim 0.5$. At small scales,
non-gravitational heating, e.g. feedback,  dominates over gravity and
leaves unique signatures in the IGM XRB, which
allows a comparable accuracy in the measurement of the amount of
non-gravitational heating and the length scales where non-gravitational
energy balances gravity.  
\end{abstract}
\section{Introduction}

   A large fraction of the intergalactic medium (IGM) is hot. The
   cosmic virial theorem 
predicts that the mean IGM temperature $>0.2$ keV \citep{Pen99}. It
emits X-ray  through thermal bremsstrahlung and contributes to the soft
(0.5-2 keV) X-ray background (XRB). About
$80$-$90\%$ of the  soft XRB has been resolved into point
objects such as AGNs 
\citep{Hasinger93}. Various sources such as nearby low luminosity AGNs
\citep{Halderson01}, unresolved galactic stars \citep{Kuntz01},
galactic gas,  X-ray sources in external galaxies and the IGM
\citep{Pen99,Wuxp01,Croft01} may contribute a significant fraction
to the remaining $10$-$20\%$.  To
distinguish those possible components, one can combine the XRB
mean flux, auto correlation \citep{Soltan94,Sliwa01} and cross
correlation with galaxies \citep{Almaini97,Refregier97}. In \S
\ref{sec:models}, we will build analytical models to estimate the IGM
contribution to the XRB flux and correlations.

In pure gravitational clustering where the only source of thermal
energy is shock heating from collapse, simulations show that the gas
correlation function $\xi_{\rm gas}$ follows the dark matter
correlation function down to scales where the gas over-density correlation
$\xi_{\rm gas}\sim10^3$ \citep{Pen99}. Non-gravitational heating (feedback)
can rearrange the gas distribution on small scales. By searching for
the scale where feedback dominates over gravity, one can robustly
measure the strength and history of non-gravitational heating. 
The IGM XRB is sensitive to small scale gas structures due to the X-ray
emissivity dependence on density squared. This makes the IGM XRB a
potentially powerful probe for studying feedback from galaxies on the
IGM. From the estimation 
of the IGM XRB flux, \citet{Pen99,Wuk01,Voit01} 
suggested that, if only gravitational  heating exists, the  soft
X-ray emission that the IGM produces would exceed the observational limit.
To suppress the  IGM clumping and reduce the X-ray emission, a
significant amount of non-gravitational injection energy $\sim 1$
keV/nucleon is required.  
\citet{Croft01,Dave01,Phillips01} argued from simulations that  such energy
injection is not  necessary.  In their simulations, a large fraction of
baryons (about $70\%$) are too cold to contribute significantly to the
XRB. Despite the stability problem of cool gas, these findings may
indicate a degeneracy between non-gravitational heating and the fraction of IGM
contributing to the XRB. The IGM XRB auto correlation function (ACF) and
cross correlation function (CCF) with galaxies have different
dependences on  the IGM thermal state and are capable of breaking  this
degeneracy. 
Furthermore, with galaxy photometric redshift data, the redshift
resolved  IGM X-ray emissivity-galaxy cross correlation and emissivity auto
correlation can be extracted. This tells the evolution of the IGM state.
In \S\ref{sec:xrbapplication}, we will forecast the sensitivity of
ROSAT all sky 
survey \citep{Voges99} and SDSS\footnote{SDSS, http://www.sdss.org/}
and  test the feasibility to constrain 
the IGM thermal history from correlations.  This estimation depends on
the XRB non-Gaussianity since it affects the error analysis. We analyze
our $512^3$ moving mesh hydro simulation to quantify this effect.

\section{Analytical models for the IGM XRB}
\label{sec:models}

The X-ray emissivity in the band $E_1\le h \nu \le E_2$ is given by:
$I_X \simeq 2.4\times 10^{-27} c_Z T^{1/2}
[\exp(-E_1/T)-\exp(-E_2/T)]  n_e^2\equiv \Lambda_{E_1}^{E_2}(T)  n_e^2\ {\rm
erg \ cm^{-3} \ s^{-1}} $
\citep{Tucker75}. $\Lambda_{E_1}^{E_2}(T)$ is the cooling coefficient
in the energy band $E_1$-$E_2$. A mean gaunt factor $1.2$ is
adopted. $c_Z\sim Z/Z_{\sun} (4 {\rm keV}/T)+1$  is a fitting formula
to result of metal cooling from \citet{Raymond76}. We will use 
$Z=\frac{1}{4}Z_{\sun}$ as the lower plausible limit on the IGM
inferred from clusters of galaxies. We choose a flat $\Lambda$CDM
cosmology with $\Omega_0=0.37$, 
 $\Omega_{\Lambda}=0.63$, $\Omega_B=0.04$, $\sigma_8=0.9$, $h=0.7$ and
$n=1$ to calculate the soft XRB statistics. Hereafter, we always
consider the comoving emissivity $I_X^C=I_X/(1+z)^3$.

The IGM XRB flux $
F_X(\hat{\theta})=\int I^C_X(\chi\hat{\theta})/[4 \pi (1+z)^2] d\chi$.
$\chi$ is the comoving distance. The mean XRB flux $\bar{F}_X= \int
\bar{I}^C_X/[4 \pi (1+z)^2] 
d\chi$. The galaxy surface  density $
\Sigma(\hat{\theta})=\int \frac{dn}{dz}[1+\delta_G(\chi\hat{\theta})]dz.$
$\frac{dn}{dz}$ is the galaxy redshift distribution and 
$\delta_G$ is the galaxy number over-density. The mean surface
density $\bar{\Sigma}=\int \frac{dn}{dz}dz$.
  We define the dimensionless fluctuations $\Delta_F\equiv
F/\bar{F}-1$ and $\Delta_{\Sigma}\equiv \Sigma/
\bar{\Sigma}-1$. The IGM XRB ACF and CCF with galaxies are defined by
$w_X(\theta)\equiv \langle \Delta_F(\hat{\theta}_1)
\Delta_F(\hat{\theta}_2)\rangle$ and $w_{XG}(\theta)
\equiv \langle \Delta_F(\hat{\theta}_1)
\Delta_{\Sigma}(\hat{\theta}_2) \rangle$, respectively. Here,
$\hat{\theta}_1 \cdot \hat{\theta}_2=\cos \theta$. $C^X(l)$ and
$C^{XG}(l)$ are the corresponding angular
power spectra, respectively.

These 2D correlations are determined by the corresponding  3D
correlations  such as the  emissivity ACF
$\xi_X(r)\equiv\langle
\delta_X({\bf x})\delta_X({\bf x+r})\rangle$ and emissivity-galaxy CCF
$\xi_{XG}(r)=\langle 
\delta_X({\bf x}) \delta_G({\bf x+r})\rangle$ or their corresponding power
spectra $P_X(k,z)$ and $P_{XG}(k,z)$. $\delta_X \equiv
I^C_X/\bar{I}^C_X-1$. At small
angular scales we use the Limber's equation to obtain
\begin{equation}
\label{eqn:clx}
C^X(l)=\frac{1}{\bar{F}_X^2}\int_0^{\chi_{\rm re}}
P_X(\frac{l}{\chi},z)\frac{[\bar{I}_X^C(z)]^2}{[4\pi(1+z)^2]^2\chi^2} d\chi, 
\end{equation}
\begin{equation}
\label{eqn:clxg}
C^{XG}(l,z_1,z_2)=\frac{1}{\bar{F}_X\bar{\Sigma}}\int_{z_1}^{z_2}
P_{XG}(\frac{l}{\chi},z)\frac{\bar{I}_X^C(z)}{4\pi(1+z)^2\chi^2}
\frac{dn}{dz} dz. 
\end{equation}
Here, $\chi_{\rm re}$ is the comoving distance to the reionization
epoch. $[z_1,z_2]$ is the redshift range of the galaxy survey
adopted. 

For these statistics, we can treat the IGM either as a continuum field
with a density and 
temperature distribution or as discrete
halos.  From these two viewpoints, we build two analytical models: the
continuum field model (\S \ref{sec:hr}) and the halo 
model (\S \ref{sec:xrbps}).

\begin{figure}
\plotone{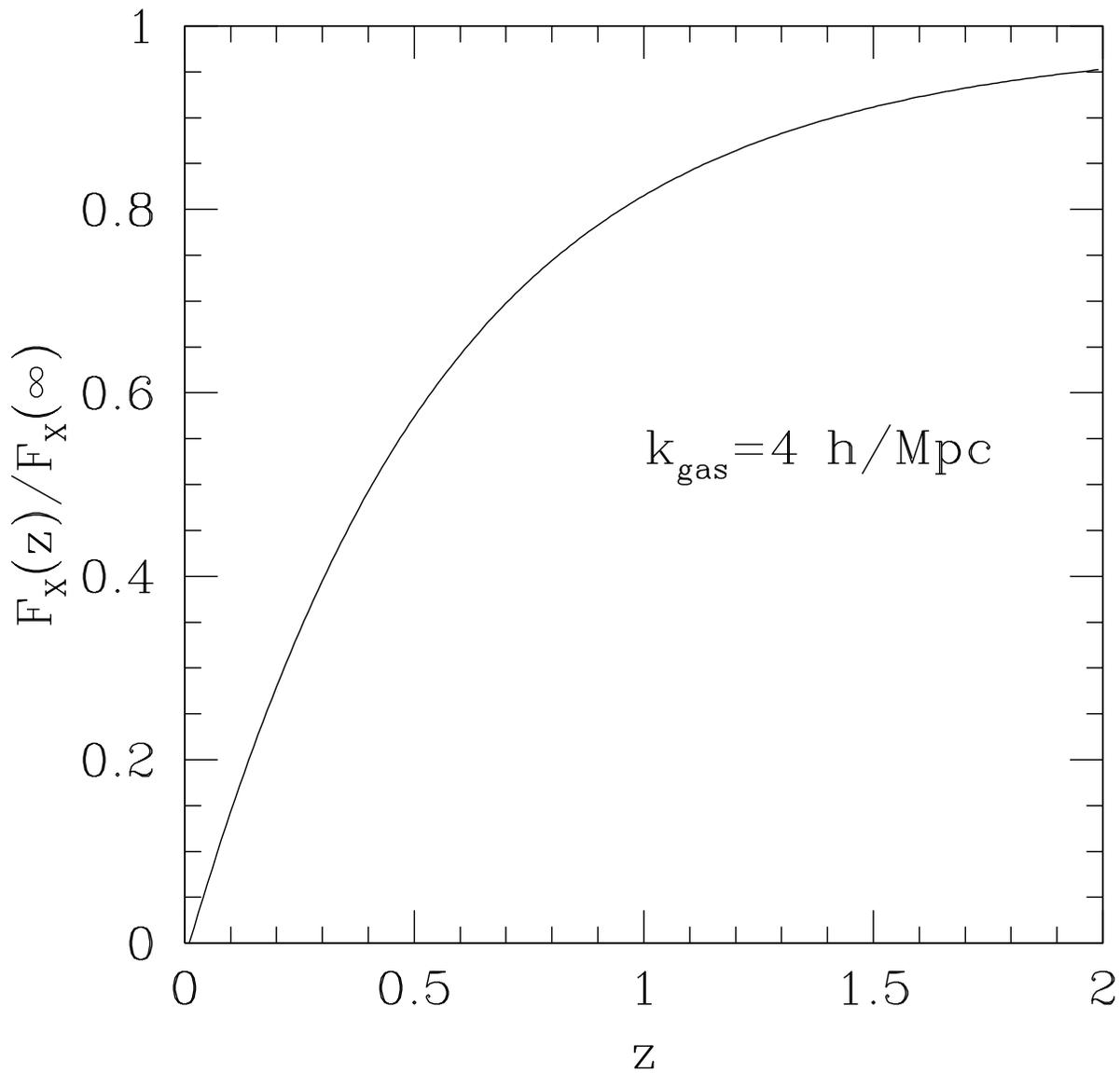}
\caption{The IGM cumulative contribution to the XRB. Contribution
with different $k_{\rm gas}$ looks similar.\label{fig:con}} 
\end{figure}

\subsection{The continuum field model}
\label{sec:hr}
For the $[0.5,2]$ keV X-ray band, we can approximate the gas
temperature $T\simeq 1$ keV due to the following arguments. (1) when
$T\ll 1$ keV, 
$\Lambda_{E_1}^{E_2}(T)$ decreases exponentially and  we would expect
that too cold gas does not 
contribute significantly to the soft XRB.  (2) When $T\gg 1 $ keV,
$\Lambda_{E_1}^{E_2}(T)$ drops as $T^{-1/2}$. Furthermore, gas with $T\gg 1
$ keV is rare. So the contribution from gas with $T\gg 1
$ keV to the soft XRB is small. (3) $\Lambda_{E_1}^{E_2}(T)$ peaks at $T\sim
1$ keV. The density weighted temperature is $\sim 0.4 $ keV
\citep{Zhang01} and the density square weighted (roughly emissivity weighted)
temperature $\sim 1 $ keV. So, we expect most contributions to the soft
XRB to be from gas with $T\sim 1$ keV.  Around this temperature, the
emissivity has only a weak temperature dependence. So, we approximate $T=1$ keV
and define a X-ray weighted gas fraction $f^X_{\rm gas}\equiv 
\sqrt{I_X/[\Lambda_{E_1}^{E_2}(1\  {\rm keV}) n_e^2]}$. From the above
argument we expect $f^X_{\rm gas}\simeq 1$, but throughout this
letter, we treat $f^X_{\rm gas}$ as a free parameter to be determined by
observations. At large scales, gas generally follows dark matter. At smaller
scales, the gas density fluctuation is suppressed by the
gas pressure. This effect can be modeled through a window function
$W_g(r)$ such that the gas over-density is a convolution of the dark
matter density and $W_g(r)$. The FWHM of $W_g(r)$ is the scale where
non-gravitational processes such as feedback begins to dominate over gravity. In Fourier space, the Fourier
component of the gas 
over-density $\delta_{\rm gas}(k)=\delta(k) W_{\rm gas}(k)$ where
$\delta(k)$ is 
the Fourier component of the dark matter over-density.
A Gaussian window function $W_{\rm gas}(k)=\exp(-k^2/k_{\rm gas}^2)$
is choose from the literature \citep{Gnedin98}. With the correlation
between gas and dark matter, one can calculate the XRB statistics using
the knowledge of dark matter density field.
\begin{figure}
\plottwo{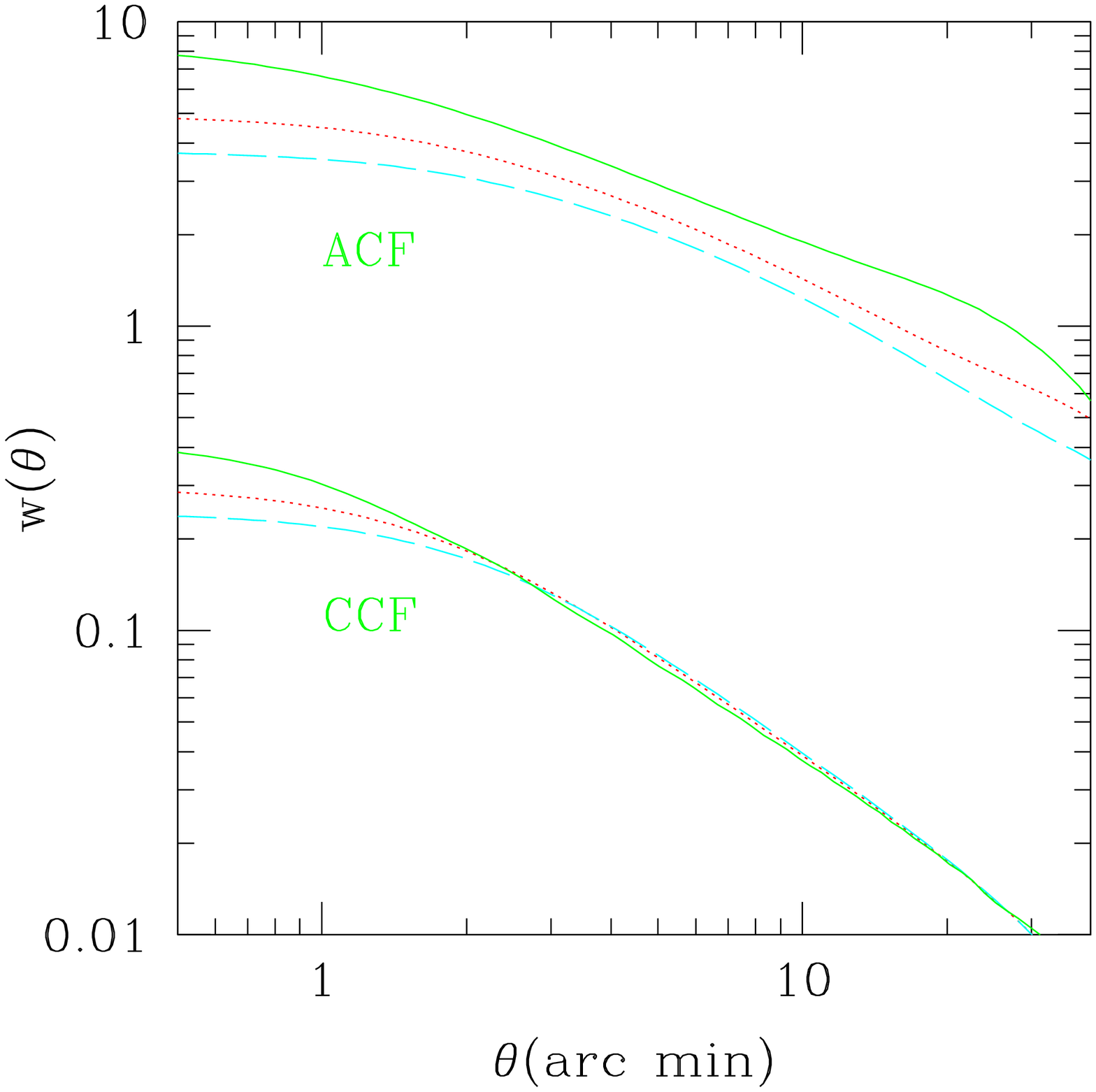}{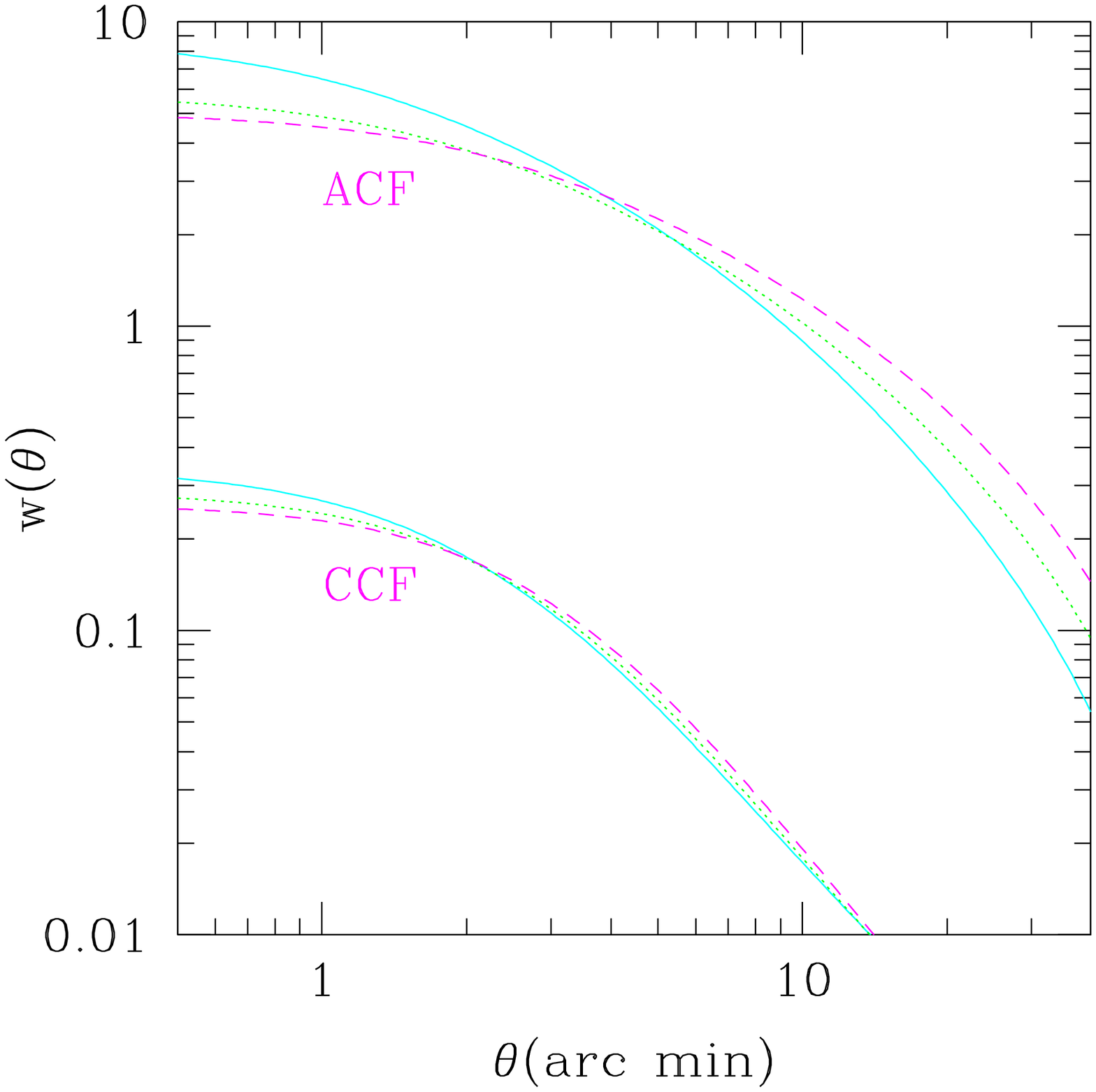}
\caption{IGM XRB angular ACF $w_X(\theta)$ and CCF $w_{XG}(\theta)$
calculated from the continuum field model (left panel) and the halo
model (right panel). The solid, dot, dash lines correspond to
$k_{\rm gas}=8,4,3 h/$Mpc ( corresponds to non-gravitational heating
energy per nucleon $E_{\rm NG}\sim 0.3,0.6,0.9$ keV) in the left panel
and $r_c(M_8,z=0)=0.5, 0.75, 1.0 h^{-1}$ 
Mpc ($E_{\rm NG}\sim 0.5,0.6,0.8$ keV) in the right panel. $k_{\rm gas}$ and
$r_c$ are the scales below which 
feedback dominates over gravity. The slopes of $w(\theta)$
predicted by  the halo model are steeper than the continuum field
results. This difference is likely due to the weak
correlation between density and temperature, which is omitted in the
continuum field model. \label{fig:acf}}  
\end{figure}

We apply the extension of the hierarchical model  \citep{Fry84} in
the highly non-linear regime: the hyper-extended perturbation theory
(HEPT) \citep{Scoccimarro99} to calculate $P_X$ and $P_{XG}$.
\begin{equation}
\label{eqn:IXC}
P_X(k,z)\simeq \frac{1}{\sigma_{\rm gas}^4 (2 \pi)^6}\int
B_4({\bf k_1},{\bf k_2},{\bf k_3},{\bf k_4};z)  W_1 W_2 W_3
W_4 d^3 k_2 d^3 k_4
\end{equation}
\begin{equation}
P_{XG}(k,z)\simeq\frac{1}{\sigma_{\rm gas}^2 (2 \pi)^3}\int b B_3({\bf
k}_1,{\bf k}_2,{\bf k};z) W_1 W_2 d^3 k_1. 
\end{equation}
Here, $W_i=W_{\rm gas}(k_i)$, ${\bf k_1}+{\bf k_2}=-{\bf k_3}-{\bf k_4}={\bf
k}$.  The gas clumping factor $\sigma_{\rm gas}^2=\int P_{\rm gas}(k)
k^2dk/(2\pi^2)$ with $P_{\rm gas}(k)$ as the gas density power
spectrum. The
bispectrum $B_3\propto P^2(k)$ and polyspectrum, the 
dominant term in the expression of $P_X(k,z)$, $B_4\propto P^3(k)$
terms are calculated from   
HEPT.  Here, $P(k)$ is the dark matter density power spectrum. See
\citet{Zhang01} for more detailed explanation. We consider a flux limited
galaxy survey SDSS,  take  the  SDSS galaxy distribution as
$\frac{dn}{dz}=\frac{3z^2}{2(z_m/1.412)^3}\exp[-(1.412z/z_m)^{3/2}]$
\citep{Baugh93} and adopt $z_m=0.43$ as the median redshift of Sloan galaxy
photometric redshift distribution \citep{Dodelson01}. We assume that
$k_{\rm gas}$ does not change with redshift and a constant bias model
$\delta_G=b \delta$ for galaxies. 

\subsection{The halo model}
\label{sec:xrbps}
The gas profile in a halo and halo mass-temperature relation
determines  the X-ray luminosity. The halo mass function  
and halo-halo correlation enable us to calculate their collective
effects to the XRB. Similar methods have been applied to the dark matter
correlation \citep{Ma00} and the Sunyaev-Zel'dovich effect
\citep{Komatsu99}.

   We adopt the electron number density  profile
$n_{e}=n_{0} \left[1+(\frac{r^{2}}{r_{c}^{2}})\right]^{-1}$.
The gas core radius $r_c$ is analogous to $k_{\rm gas}$ and corresponds to the
scale below which 
feedback significantly changes gravitational clustering. We assume 
that gas accounts for $\Omega_B/\Omega_0=11\%$ of the halo mass. The
gas temperature is determined by the virial theorem through
the relation $ 
\frac{M}{M_{8}}=\left[\frac{T/{\rm keV}}{4.9
\Omega_0^{2/3}\Omega(z)^{0.283} (1+z)}\right]^{3/2}$ \citep{Pen98}. 
$M_8$ is the mean mass contained in a $8h^{-1}{\rm Mpc}$ sphere
today. The
distribution of halo comoving number density $n$ as a function of halo
mass $M$ and $z$ is given by \citep{Press74}:
$\frac{dn}{dM}=(\frac{2}{\pi})^{1/2}\frac{\rho_{0}}{M^{2}}\frac{\delta_{c}}{\sigma}|\frac{d\rm{ln}\sigma}{d{\rm{ln}}M}|\exp(-\frac{\delta_{c}^{2}}{2\sigma^{2}})$.  
Here $\rho_{0}$ is the present mean matter density of the
universe. $\sigma(M,z)$ is the linear theory rms density fluctuation
in a sphere containing mass M at redshift $z$.  $\delta_{c}$ is
the linearly extrapolated over-density at which  an object virializes.
Its dependence on cosmology is weak and we will adopt the value
$1.686$, which is the $\delta_c$ for a $\Omega_0=1.0$ CDM universe
\citet{Eke96}. 
\citet{Mo96} related the halo-halo correlation $P_c(M_1,M_2)$ with the
underlying dark matter correlation $P(k)$ by a linear bias:
$P_c(k,M_1,M_2)=P(k) b(M_1) b(M_2)$. We adopt the NFW profile
\citep{Navarro96} with a compact parameter $c=5$ to
calculate the XRB-dark matter cross correlation. 

   In this model, the variance of the gas density $\sigma_{\rm
   gas}^2\propto \int_{M_{\rm low}}^{\infty} 
   n_0^2r_c^3[\tan^{-1}(r_{\Delta}/r_c)-\frac{r_{\Delta}/r_c}{1+(r_{\Delta}/r_c)^2}]\frac{dn}{dM}dM$.
   $r_{\Delta}$  is the virial radius and we calculate it using the
   fitting formula of \citet{Eke96}.
   Defining $\delta_X({\bf k},M)$ and $\delta({\bf k},M)$ as the
Fourier transform of $I^C_X({\bf r},M)/\bar{I}^C_X$ and 
$\delta({\bf r},M)\equiv \rho_{DM}({\bf r},M)/\bar{\rho}_{DM}$ of each halo, 
respectively, we obtain  

\begin{equation}
\label{eqn:pspx}
P_X(k)=\frac{1}{\bar{I}_X^2}\left(\int_{M_{\rm low}}^{\infty} \delta_X^2(k)
\frac{dn}{dM}dM+P(k)\left[\int_{M_{\rm low}}^{\infty} 
\delta_X(k)\frac{dn}{dM}dM b(M)\right]^2 \right),
\end{equation}
\begin{eqnarray}
\label{eqn:pspxg}
P_{X\delta}(k)&=&\frac{1}{\bar{I}_X}\left(
\int_{M_{\rm low}}^{\infty} \delta_X(k)\delta(k)
\frac{dn}{dM}dM +\right. \\ \nonumber
&&\left. P(k)\left[\int_{M_{\rm low}}^{\infty} 
\delta_X(k)\frac{dn}{dM}dM b(M)\right]\left[\int_{M_{\rm low}}^{\infty}
\delta(k)\frac{dn}{dM}dM b(M)\right] \right).
\end{eqnarray}
The integrals in these equations depend strongly  on the halo lower
mass limit  $M_{\rm low}$, which can not be arbitrarily chosen. A smaller
$M_{\rm low}$ will produce a bigger gas
clumping factor $\sigma^2_{\rm gas}$ since more gas contributes. It
will also produce a
smaller $P_X$ since gas in less massive halos is more diffuse and 
contributes a smaller fraction to the correlation than to the mean
flux. This behavior contradicts the $P_{X} \propto 
\sigma^2_{\rm gas}$ behavior we should expect.  Thus $M_{\rm low}$ must
be determined independently.  Here we adopt the model of
\citet{Pen99}.  In this model, the energy injection from feedback such
as supernovae winds expand the gas and produces a core (radius $r_1(M)$) with a
constant entropy. Then $r_1(M)=r_{\Delta}(M)$ sets the value of
$M_{\rm low}$. In this
model, $r_1\simeq 0.5 r_c\propto T^{-1}\propto M^{-2/3}$. 
 We further assume $r_c(M_8,z)=r_c(M_8,z=0)/(1+z)$, which corresponds to a
redshift  independent $k_{\rm gas}$.

\subsection{Predictions}
Two models give consistent predictions for the XRB. More than $80\%$
of the 
contribution to the soft XRB flux is from the IGM at $z\leq 1$
(fig. \ref{fig:con}). The mean X-ray 
flux $\bar{F}_X\simeq (\frac{\sigma_g^2}{30}) 10^{-12} \ {\rm erg} \ {\rm s}^{-1} \ {\rm
cm}^{-2} \ {\rm deg}^{-2}$, which accounts for a significant, if not dominant,
fraction of the unresolved XRB.   The IGM XRB is homogeneous with a
large amplitude (fig. \ref{fig:acf}) and is sufficient to
explain the observed XRB correlations. The point-to-point XRB variance
$W_X(0)\sim
10$ and $W_{XG}(0)\sim 0.3$. For  $\theta>1^{'}$, we find that $W_{XG}\propto
\theta^{\alpha}$ ($\alpha\sim -1.1$).  The shape and amplitude of these
properties have different 
dependences on gas parameters.  (1) For the shapes, the larger
$k_{\rm gas}$ (corresponds to a smaller $r_c$  and a larger
$\sigma_{\rm gas}^2$), the steeper the 
correlations. (2) For the amplitudes, $\bar{F}_X \propto 
(f^X_{\rm gas})^2 \sigma_{\rm gas}^2$ while $P_X(k),P_{XG}(k),w_X(0),
w_{XG}(0)\propto (f^X_{\rm gas})^0 \sigma_{\rm gas}^2$. Combining correlation data and mean flux data, one can 
distinguish $f_{\rm gas}^X$ from $\sigma_{\rm gas}^2$. With redshift
resolved $P_X(k,z)$ or $P_{XG}(k,z)$, as can be obtained from XRB
observations and galaxy surveys, one can infer $\sigma_{\rm gas}^2(z)$
and then $k_{\rm gas}(z)$ and $r_c(z)$ from their relations with
$\sigma_{\rm gas}^2(z)$. The interpretation of these quantities depends
on processes changing the gas-dark matter correlation such as
non-gravitational heating and possibly gravitational shock-heating. 
Current simulations have difficulties to resolve these small scales
where these processes become dominant due to limited  simulation mass
and spatial resolution. Simulation results about the IGM XRB flux, hot gas
fraction, etc. have not converged, so hereafter we focus on the
estimation of the effect of 
feedback through our analytical models and use our hydro simulation to
calibrate some XRB statistics. Feedback increases the gas 
temperature and thus $f_{\rm gas}^X$, the fraction of gas contributing to
XRB. But it dilutes the  gas and reduces 
$\sigma_{\rm gas}^2$, resulting in a larger $r_c$ or a smaller
$k_{\rm gas}$. From Pen's model, for a $\Lambda$CDM universe, we can
estimate $E_{\rm NG}(z)$, the  non-gravitational energy injection per nucleon
in unit of  keV, from the relation $\sigma_{\rm
gas}^2 \simeq 90 \Omega_0[2\sqrt{E_{NG}}\exp(-E_{\rm NG})+\sqrt{\pi} {\rm
erfc}(\sqrt{E_{\rm NG}})]/E_{\rm NG})$.   $k_{\rm
gas}(z)$ (or $r_c$)  and $E_{\rm NG}(z)$  then tell us the scales at
which feedback dominates over gravity and the feedback strength.

\begin{figure}
\plotone{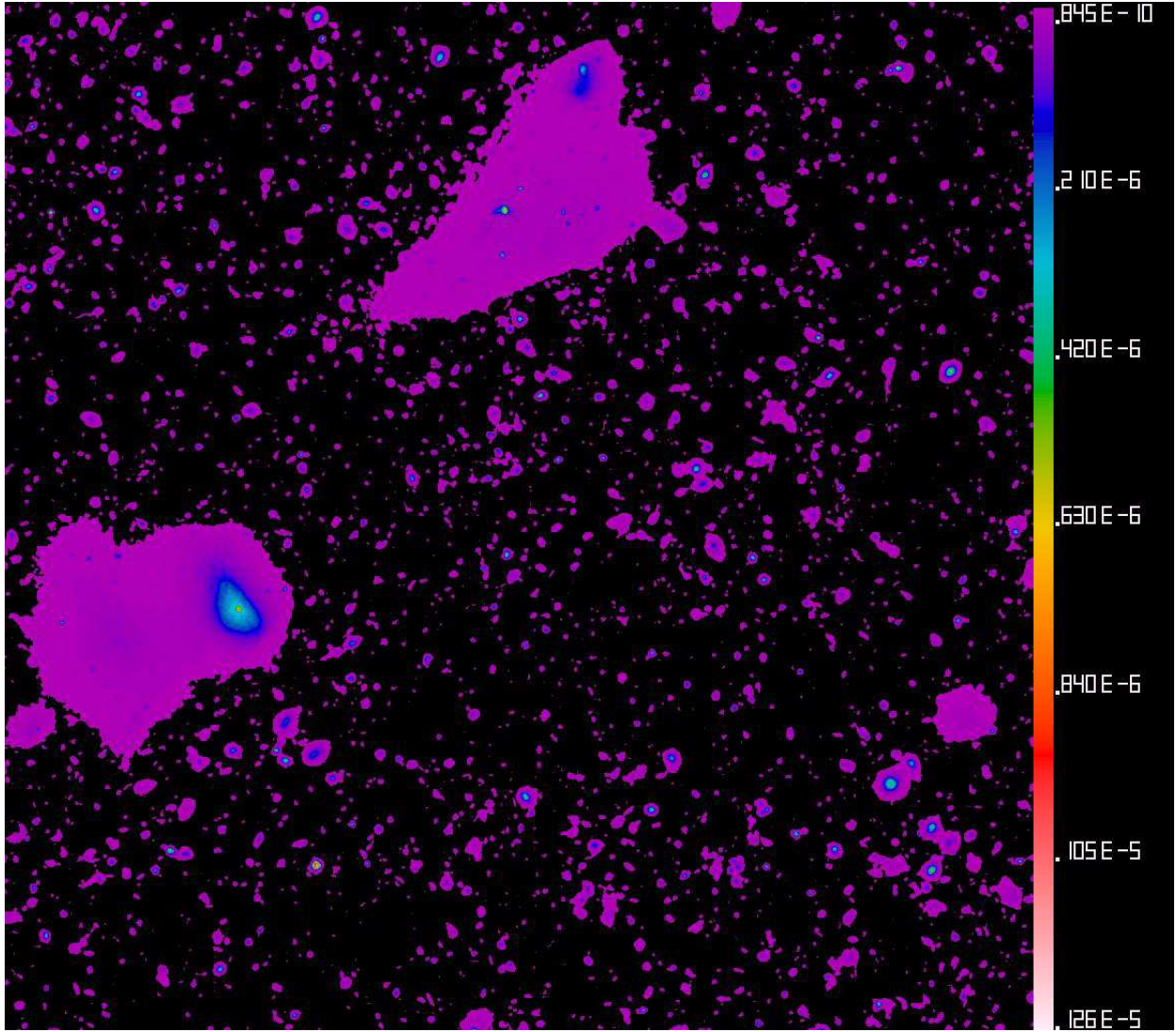}
\caption[A XRB map in our $512^3$ MMH simulation]{A XRB map in our $512^3$ MMH
  simulation\citep{Zhang02}. The parameters we adopted in this simulation are
$\Omega_0=0.37$, $\Omega_{\Lambda}=0.63$, $\Omega_B=0.05$, $h=0.7$,
$\sigma_8=1.0$, the primordial power spectrum index $n=1$, box size
  $L=100 h^{-1}$ Mpc and smallest grid spacing $40 h^{-1}$ kpc. The
  XRB map integrates contributions up to $z=3$ and has a width of $1.32$
  degree. The XRB flux is in units of $  
  {\rm erg} \ {\rm s}^{-1} \ {\rm  cm}^{-2} \ {\rm
  sr}^{-1}$. \label{fig:xrbmap}} 
\end{figure}

\begin{figure}
\plotone{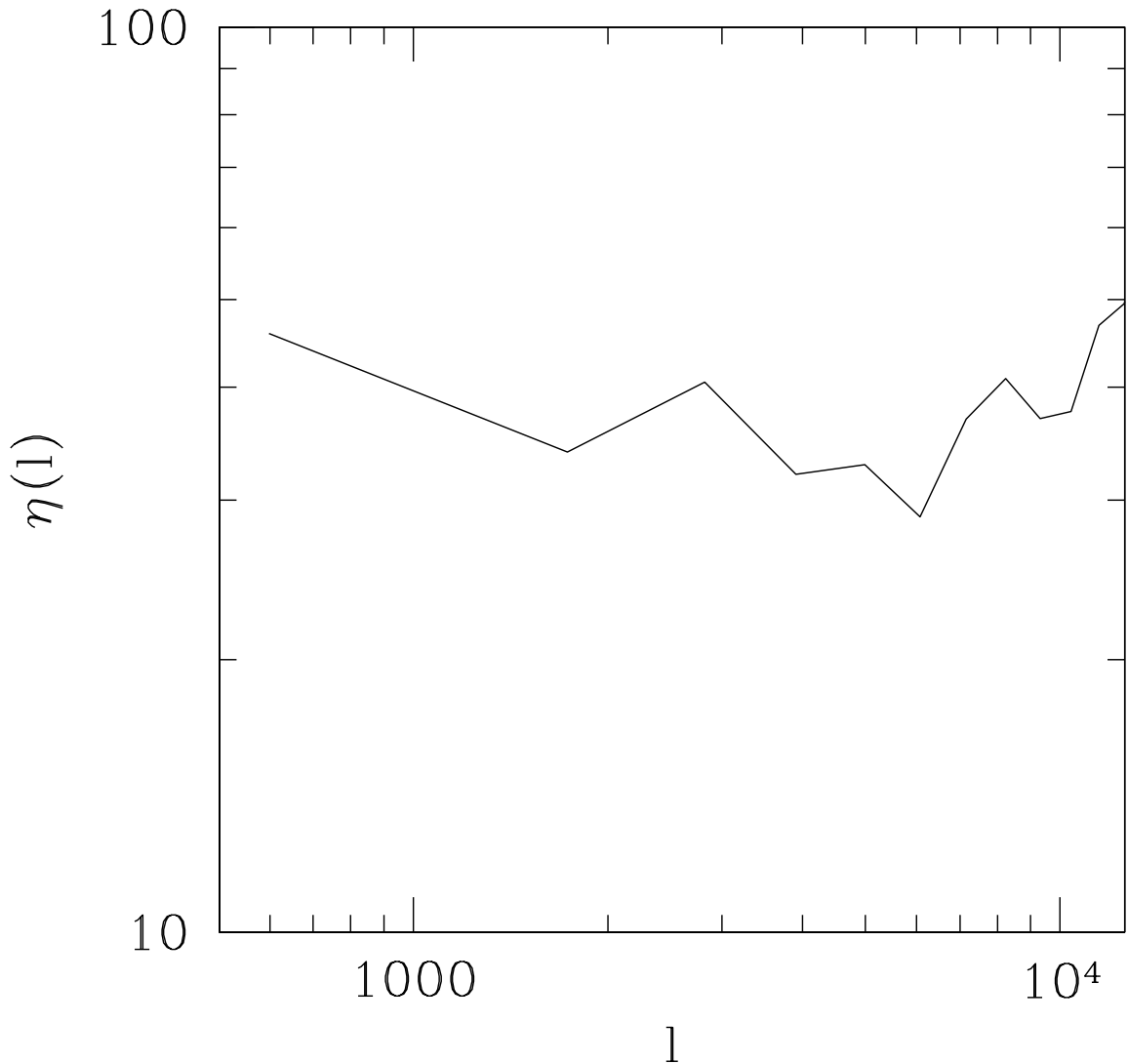}
\caption[The XRB non-Gaussianity]{The XRB-galaxy cross correlation
non-Gaussian corrections $\eta(l)$ from sample variance errors
measured from our 
$512^3$ MMH simulation. For Gaussian signals, $\eta(l)=0$. We make
$40$ XRB maps and $40$ maps of the intergalactic gas 
surface density at the same fields of view to measure the XRB-galaxy
cross correlation assuming 
that galaxy number density traces the gas distribution. We then
measure the fluctuation of the cross correlation power spectra of
these maps to obtain the XRB-galaxy cross correlation power spectrum
non-Gaussianity $\eta(l)$.  We find a strong non-Gaussian correction
at scales $500<l<10^4$. \label{fig:XRBngs}}
\end{figure}

\begin{figure}
\plottwo{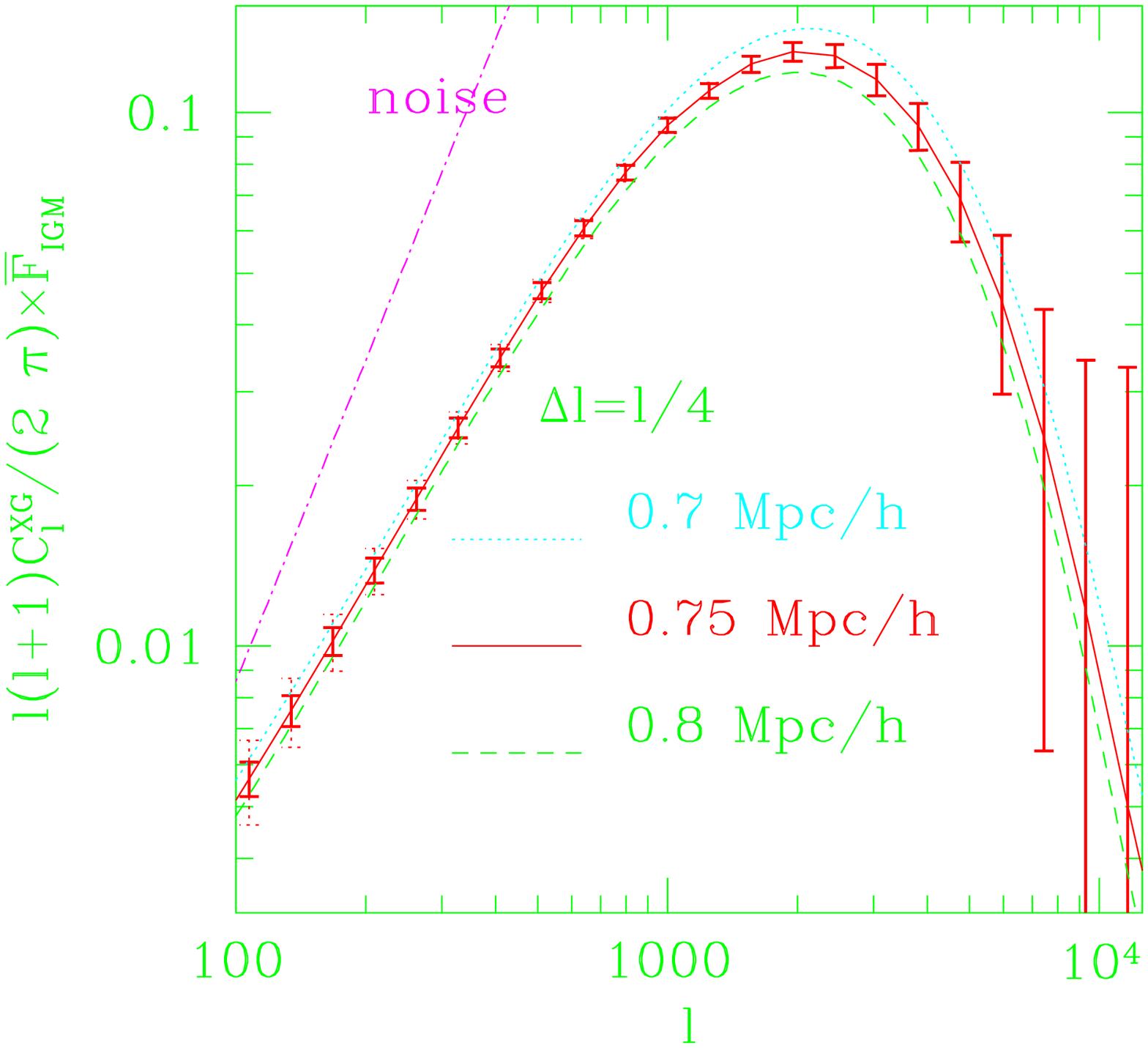}{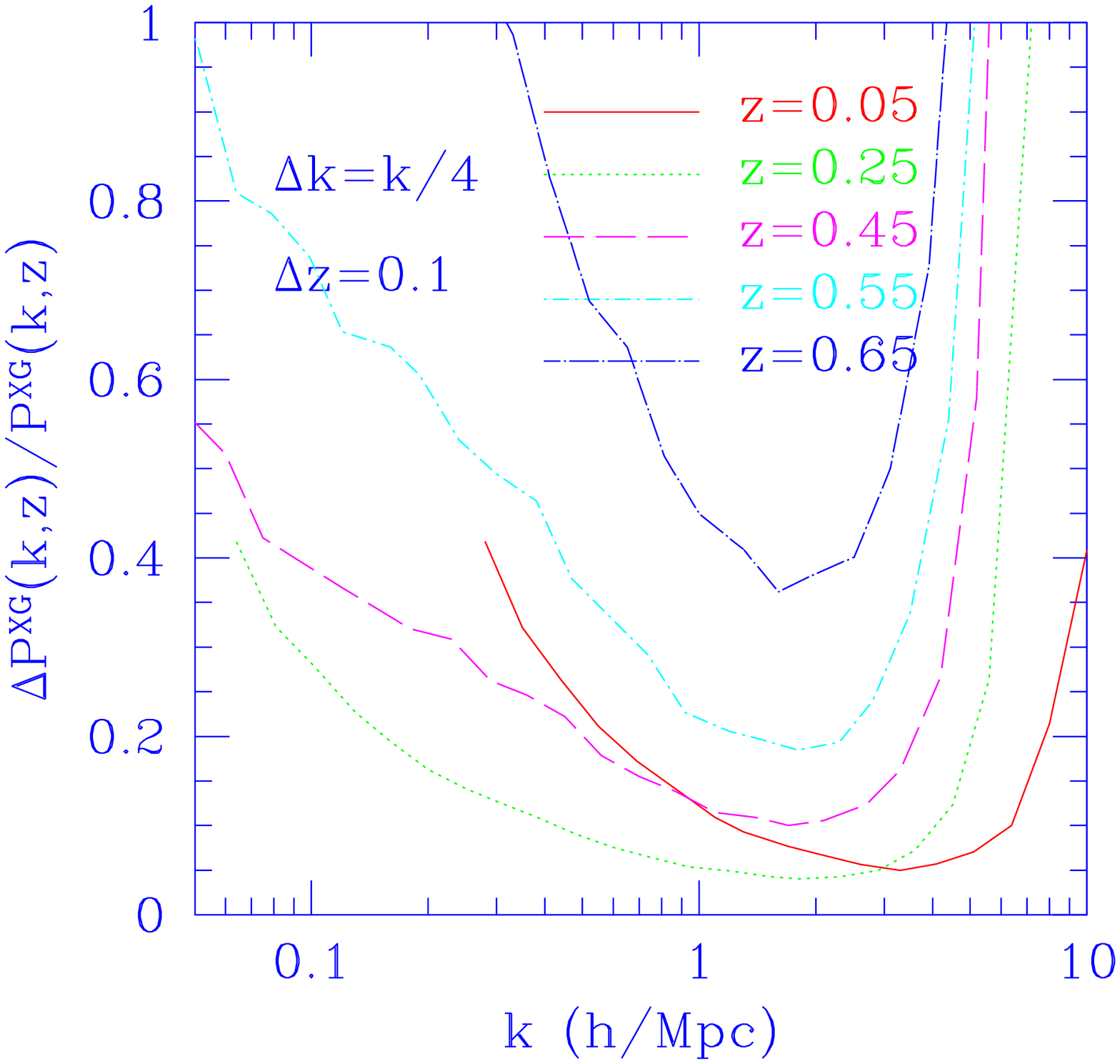}
\caption{Statistical errors in the forecast ROSAT+SDSS measurement of
$\frac{l(l+1)}{2\pi} C^{XG}(l)\bar{F}^{\rm IGM}_X$ (left panel) and
extracted $P_{XG}(k,z)$ (right panel). In the left panel, error bars
are for the central
line with $r_c(M_8,z=0)=0.75h^{-1}$ Mpc and we normalize its
$\bar{F}^{\rm IGM}_X=1$. Solid error bars are Gaussian errors and
the dot error bars are non-Gaussian errors. The effect of
non-Gaussianity is dominant at large angular scales and is negligible
at small angular scales where the dominant error is from noise.  The redshift
averaged gas parameter $r_c$  
can be constrained with $\sim 10\%$  accuracy at $1\sigma$. The amount
of feedback and the scale below which feedback 
dominates over gravity can be modeled with a comparable accuracy as 
$P_{XG}(k,z)$.\label{fig:xrberror}}  
\end{figure}

\section{Extracting the IGM state and evolution}
\label{sec:xrbapplication}
X-ray sources and differential extinction of our galaxy make the
measurement of the IGM XRB flux and 
ACF difficult. But the measurement of IGM XRB-galaxy CCF is 
much more robust. The direct observable in the CCF measurement is   
$\langle F_X(\hat{\theta}_1) \Sigma_G(\hat{\theta}_2)\rangle$.
We need to estimate the IGM contribution to the above property in
order to infer $w_{XG}^{\rm IGM}(\theta)$. (1) Our
calculation suggests that  the IGM XRB is sufficient to explain the
unresolved X-ray flux and the cross correlation with galaxies.
(2) Distant AGNs and galactic X-ray sources have almost no
correlation with nearby galaxies with $z\lesssim 1$. (3) The CCF between
galaxies and X-ray sources in extragalactic galaxies or nearby low
luminosity AGNs is of 
the same amplitude of galaxy surface density ACF, which is one order
of magnitude lower than  the IGM XRB-galaxy CCF. So, even if they
contribute a comparable amount to the XRB flux as the IGM, their
contribution to the cross correlation is negligible. (4) For a low matter
density universe, the CCF caused by the weak lensing of low redshift
large scale structures to unresolved high redshift AGNs accounts for
at most $1/3$ of the correlation  \citep{Cooray99}.  In principle,
combining  the XRB mean flux, auto correlation and cross correlation
measurement, the IGM XRB cross correlation can be determined. We
neglect possible systematic errors in such measurements and  take the ROSAT
all sky survey and SDSS as our  targets to estimate the statistical
error in  the IGM XRB CCF measurement.

The ROSAT all sky survey (RASS) covers the whole sky ($f^{\rm RASS}_{\rm
  sky}\simeq 1$) in the $0.1$-$2.4$ soft X-ray
band in $t=1.03\times 10^7$ s. At the energy bands
below $0.5$ keV,  the galactic emission and absorption are strong and
at bands $>1$ keV,  only a fraction of the sky can be used. So, we
choose the band $0.5$-$1$ keV for our analysis. A factor of $4$
  decrease in the photon count rate is expected due to  the choice of  this
  band instead of the whole RASS band\footnote{The
  factor of $4$ is communicated to us by the referee Andrzej Soltan.}.  
One full field
view of RASS has a $\sim 1$ degree radius and only within the central
$10$-$15$ arc-min radius is the resolution better than $3$ arc-min. Since
we need high resolution to probe the IGM state, we will focus on these
central regions. Since the separation of two successive scans is
smaller than $10$ arc-min, these high resolution regions cover almost the
whole sky. But the number of photons received in these regions is
about $1/36$-$1/16$th of the total number of photons received.  We adopt
the conservative fraction $1/36$ for our following estimation. The
  SDSS covers   
$f^{\rm SDSS}_{\rm sky} \simeq1/4$ fraction of the sky and will detect about
$N_G=5\times 10^7$ galaxies.  We estimate the error in 
the power spectrum $C_l$ measurement. The error in the IGM XRB-galaxy power spectrum is 
\begin{equation}
\label{eqn:clxgerr}
\Delta C_l^{XG}=\sqrt{\frac{(\eta(l)+1)[C^{XG}_l]^2+(C^X_l+C_N^X) \times (C^G_l+C_N^G)}{(2l+1)\Delta l
f^{\rm SDSS}_{\rm sky}}}.
\end{equation}
$C^G$ is the power spectrum of the galaxy surface
density. $C_N^G(l)=4\pi f^{\rm SDSS}_{\rm sky}/N_G$ is the Poisson noise of the
galaxy number  
count. 
$C_N^X(l)=4\pi f^{\rm RASS}_{\rm
sky}(\bar{F}_N/\bar{F}_X)^2/N_N+4 
\pi f^{\rm RASS}_{\rm sky}/N_X$, where the first term is the
Poisson noise  of 
the RASS background and the second term is the Poisson noise of the
IGM XRB signal.  $N_N$ and $N_X$ are the total numbers of photons
that the RASS received in the band $0.5$-$1$ keV from background noise
and the IGM, respectively. We adopt $\bar{F}_N(0.1$-$2.4$ keV$)\simeq
5\times 10^{-12} {\rm 
  erg} \ {\rm cm}^{-2}\ {\rm s}^{-1} {\rm deg}^{-2}$ (Fig. 3,
\citet{Voges99}). 
$\Delta l$ is the bin width and $f_{\rm
sky}^{\rm SDSS}$ reflects the cosmic variance. The strong
non-Gaussianity of the XRB affects its error analysis. One could in practice 
estimate
$\eta(l)$, the non-Gaussianity of the XRB-galaxy cross correlation
power spectrum (for Gaussian case, $\eta=0$) from
our models.  For simplicity, we will adopt the result  from our
$512^3$ moving mesh hydro (MMH) simulation \citep{Zhang02}.  The
parameters we adopted in our $512^3$ simulation are 
$\Omega_0=0.37$, $\Omega_{\Lambda}=0.63$, $\Omega_B=0.05$, $h=0.7$,
$\sigma_8=1.0$, power spectrum index $n=1$, box size $L=100 h^{-1}$
Mpc and smallest grid spacing $40 h^{-1}$ kpc.  Though the cosmology
adopted in this simulation is not identical to the fiducial cosmology
we adopt in this paper, it should yield a rough estimate of the XRB
non-Gaussianity. During this adiabatic simulation we
store 2D projection of the X-ray flux through the 3D box at every
light crossing time  through the box. The projections are made
alternatively in the x, y, z  directions to minimize the repetition of
the same structures in the projection. Our
2D maps are stored on $2048^2$ grids. After the simulation, we stack
the XRB sectional maps 
separated by the width of simulation box, randomly choosing the center
of each section and randomly rotating and flipping each section.  The
periodic boundary condition guarantees that there are no
discontinuities in any of the maps. We then add these sections onto a
map of constant angular size.  Using different random seeds for the
alignments and rotations, we make $40$ maps of width $1.32$
degrees by integrating from zero to  $z=3$ (Fig. \ref{fig:xrbmap}). Using the same set
of random seeds, we make $40$ maps of intergalactic gas
surface density to measure the XRB-galaxy cross correlation assuming
that galaxy number density traces the gas distribution. We then
measure the fluctuations of the cross correlation power spectra of
these maps to obtain the XRB-galaxy cross correlation power spectrum
non-Gaussianity, namely, $\eta(l)$ (figure 
\ref{fig:XRBngs}). The details of this simulation are described in
\citet{Zhang02}. The error 
is dominated by this non-Gaussian cosmic 
variance  at large angular scales and by noise at small angular
scales (fig. \ref{fig:xrberror}). $C^{XG}(l)$ can be
measured to a better than $10\%$ accuracy for $800<l<5000$. If we
cross correlate galaxies at 
redshift bin $[z_i-\Delta z/2,z_i+\Delta z/2]$ with the XRB and if $\Delta
z$ is sufficiently small, 
$C^{XG}(l,z_i-\Delta z/2,z_i+\Delta z/2)\rightarrow
P_{XG}(\frac{l}{\chi(z_i)},z_i) \chi^{-2}(z_i)/\Delta \chi$
(eqn. \ref{eqn:clxg}). This 
equation enables one to infer the redshift resolved $P_{XG}(k,z)$. The
error in the $P_{XG}(k,z)$ estimation is given by
eqn. (\ref{eqn:clxgerr}) with all 
$C_l$ being replaced by $C_l(z_i-\Delta z/2,z_i+\Delta z/2)$ except for
$C_N^X$, where $\bar{F}_N$ should remain unchanged due to the absence of
redshift 
information. We choose $\Delta z=0.1$.  $P_{XG}(k,z)$ can be extracted up to
$z\sim0.6$ and the error around the peak of $\Delta^2_{XG}(k)\equiv
\frac{k^3}{2\pi^2}P_{XG}(k)$ is $\lesssim 20\%$
(Fig. \ref{fig:xrberror}). Around this peak the dominant source of
errors is noise and the effect of non-Gaussianity is negligible, the
uncertainty of the $\eta(l)$ estimation does not affect the accuracy
of $P_{XG}(k,z)$ around its peak and the subsequent $\sigma^2(z)$
extraction. The cross correlation coefficient $r(k,z)\equiv 
P_{XG}(k,z)/\sqrt{P_X(k,z)P_G(k,z)}$ has a weak dependence on
$z$ and enables one to infer the redshift
dependence of $P_X(k,z)$ from the measurement of $P_{XG}$ and
$P_G$. Given this redshift dependence, one can invert the observable
two-dimensional angular power spectrum $C^X(l)$ to a three-dimensional
power spectrum $P_X(k,z)$. 

From these measurements, the feedback history can
be extracted. The redshift averaged $r_c$ could be determined with
a $10\%$ accuracy (fig. \ref{fig:xrberror}). $\sigma^2_{\rm
gas}(z)\propto P_{XG}(z)$ could be determined 
with $\sim 20\%$ accuracy for $z\lesssim 0.5$
(fig. \ref{fig:xrberror}). Since $\frac{\delta
\sigma_{\rm gas}^2}{\sigma_{\rm gas}^2}\sim \frac{\delta 
k_{\rm gas}}{k_{\rm gas}} \sim \frac{\delta r_c}{r_c} \sim \frac{\delta E_{\rm
NG}}{E_{\rm NG}}$, the feedback amount $E_{\rm
NG}$ and the scale $k_{\rm gas}$ or 
$r_c$, at which  feedback dominates over gravity,  can in principle be
extracted with a comparable accuracy. Its calibration would require
simulations with feedback, which are currently being studied.  

 Our estimation shows the sensitivity of the XRB statistics to the gas
profile. In our estimates, the gas profile is taken as a free function and the
gas fraction is taken to be the same for each halo. In practice, feedback
changes both  the gas profile and the gas fraction.  This complication
does not affect our estimation of the
feasibility to extract the gas state from XRB observations, since it
only depends on the sensitivity of the XRB statistics to the gas
profile. But it
does affect the interpretation of the data, for example, the relation
between the XRB statistics and the amount of feedback. A further
investigation of this issue requires a detailed study of the gas
profile and its evolution when feedback presents. We are currently
carrying out a self-consistent calculation of the effect of the
feedback on the XRB, the thermal Sunyaev Zel'dovich effect and the
cluster $L_X$-$T$ relation and will check it in hydro simulations
\footnote{Zhang, Pengjie \& Pen, Ue-Li, 2003, in preparation}.

\section{Conclusion}
\label{sec:xrbconclusion}
We probed the feasibility to extract the IGM state and evolution from
the combination of XRB surveys such as RASS and galaxy surveys such
as SDSS. To do that, we first built two analytical models, the
continuum field model and the halo 
model to calculate the statistics of the IGM XRB. The two models give
consistent results on the IGM XRB flux and correlations.  We found
that the IGM may contribute a  significant,  
if not dominant, fraction to the unresolved soft XRB 
flux, its auto correlation and cross correlation with galaxies. Since
these statistics have different dependences on gas parameters such as
the hot gas fraction and gas clumping factor, we
suggest that by combining the XRB flux and correlation observations, hot
gas fraction and hot gas clumping factor could be extracted
simultaneously. At small 
 scales, non-gravitational heating such as feedback from galaxies
dominates over gravity. This changes the gas power spectrum
and leaves signatures in the IGM XRB statistics and allows its
extraction from XRB observations. From our models and the hydro
simulation calibrated error estimation, we estimated that
RASS+SDSS would constrain the gas clumping factor to a
better than $20\%$ accuracy up to $z\lesssim 0.5$. The amount
of feedback and the scales where feedback dominates over gravity can
be extracted with a comparable accuracy. 

{\it Acknowledgments:} We thank Andrzej Soltan and Wolfgang Voges for
detailed discussions of the RASS.

\end{document}